# The vibration research of the AC dipole-girder system for CSNS/RCS

Liu Ren-Hong (刘仁洪)[1,2;1]  Zhang Jun-Song(张俊嵩)[1]  Qu Hua-Min (屈化民)[1]  Kang Ling (康玲)[1]

Wang Mo-Tuo(王莫托)[1]  wang Guang-Yuan (王广源)[1]  Wang Hai-Jing (王海静)[1,2]

[1] Institute of High Energy Physics, Chinese Academy of Sciences, Beijing 100049, China

[2] University of Chinese Academy of Sciences, Beijing 100049, China

**Abstract:** China Spallation Neutron Source (CSNS) is a high intensity proton accelerator based facility, and its accelerator complex includes two main parts: an H- linac and a Rapid Cycling Synchrotron (RCS). The RCS accumulates the 80MeV proton beam, and accelerates it to 1.6GeV, with a repetition rate of 25Hz. The AC dipole of the CSNS/RCS is operated at a 25 Hz sinusoidal alternating current which causes severe vibration. The vibration will influence the long-term safety and reliable operation of the magnet. The dipole magnet of CSNS/RCS is active vibration equipment which is different with ground vibration accelerator. It's very important to design and research the dynamic characteristic of the dipole-girder system. This paper takes the AC dipole and girder as a specific model system, a method for researching the dynamic characteristic of the system is put forward by combining theoretical calculation with experimental testing. The ANSYS simulation method plays a very important role in the girder structure design stage. With the method the mechanical resonance phenomenon was avoided in the girder design time. At the same time the dipole vibratory force will influence the other equipment through the girder. It's necessary to isolate and decrease the dipole vibration. So a new isolator was designed to isolate the vibratory force and decrease the vibration amplitude of the magnet.

## 1 Introduction

The CSNS-I accelerators consist of an 80-MeV H− linac and a rapid cycling synchrotron of 1.6 GeV[1-2]. The RCS ring is a four-folded symmetrical topological structure which consists of four arc zones and four line segments. There are 24 sets dipole magnets uniformly distributed in the whole RCS ring, and the magnets will be operated at a 25 Hz rate sinusoidal alternating current. The magnetic core and coils will make severe vibration especially at the frequency 25 Hz such as the J-PARC AC dipole magnets. The vibration influenced other equipment through the magnet girder system.

The AC dipole-girder system with complex structure and high-precision adjustment is one of the most important equipment of the CSNS/RCS. Because of the self-excited vibration, the comprehensive technical index of requirement is different from other accelerators which vibration was caused by the ground vibration. So it is necessary to study the dynamic characteristic and reduce the vibration of the system [3-4]. The theoretical modal analysis and testing modal analysis are the main research methods. The theoretical modal analysis is based on the liner vibration theory and finite element method to research the relationship among the excitation, system and response. Domestic and foreign scholars obtained many achievements by the theoretical modal analysis. The testing modal analysis uses the input and response parameters to obtain the modal parameters (frequency, damping ratio and vibration mode) [5]. The dynamic characteristic of the girder is very important. This paper adopts the AC dipole-girder system as the research object. The theoretical and testing methods are used to study the dynamic characteristic of the system. After that a new isolator was designed to improve the dynamic characteristic of the system, decrease the vibratory force and avoid the resonance phenomenon.

## 2 Vibration testing of the AC dipole

The CSNS/RCS old dipole magnet is placed on the magnetic measurement girder, and the dipole was operated at a 25 Hz sinusoidal alternating current of 1100 DC with 816 AC which causes severe vibration. The acceleration sensor will be used to measure vibration of the dipole，and the testing project and results are shown in

---

1) Email: liurh@ihep.ac.cn

Fig. 1. The maximum amplitude is 7.16 um at the vertical direction(Y). The main amplitude is at 25 Hz which is 6.31 um, and the frequency doubling of 50 Hz and 75 Hz are much less than exciting frequency. At the same time the vibration influenced the other equipment through the magnetic measurement girder. The iron core and coil of the old CSNS/RCS dipole taken crack phenomenon in the magnetic measurement phase. And the sixth natural frequency of the dipole-magnetic measurement girder system is 24.75 Hz which is quite closed to the exciting frequency 25 Hz [6]. The sixth natural frequency maybe amplifies the vibration amplitude of the dipole. So the final girder of the AC dipole must avoid resonance phenomenon through structure design or use vibration isolator.

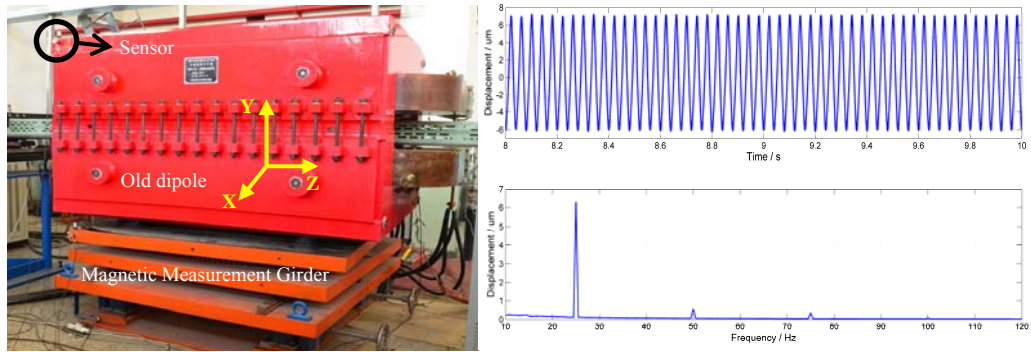

Fig.1. The old dipole vibration testing of the CSNS/RCS

## 3 Modal analysis of the dipole-girder system

### 3.1 Theoretical modal analysis simulation

The system suffers from the vibrating force which comes from the magnet. The whole structure is a multi-degree-of-freedom system, and the vibration differential equation can be expressed in the formula below,

$$MX'' + CX' + KX = F(t). \quad (1)$$

where $M$ is the system mass matrix, $C$ and $K$ are the damping and stiffness matrix, $X''$ is the system acceleration matrix, $X'$ and $X$ are the velocity and displacement matrix. $F(t)$ is the vibrating force matrix of dipole magnet.

The response of the whole system can be regard as the superposition of the natural frequency and vibration mode parameters in the state of non-damping free vibration. The nonzero solution condition of the constant-coefficient-linear-homogeneous differential of Eq. (1) is

$$|K - \omega^2 M| = 0. \quad (2)$$

The natural frequency $\omega_i^2$ and main vibration mode $\{\varphi_i\}$ can be obtained from Eq. (2), where $i$=1, 2,…, n. The finite element calculation is the main method to simulation the dynamic characteristic of the structure. This paper uses ANSYS software to simulation the AC dipole-girder system of CSNS/RCS. The dipole girder was designed with split type structure to increase the stiffness and decrease the adjustment coupling of the girder. The main structure of the girder is reinforcing steel plate and three direction adjustment system which are shown in Fig. 2.

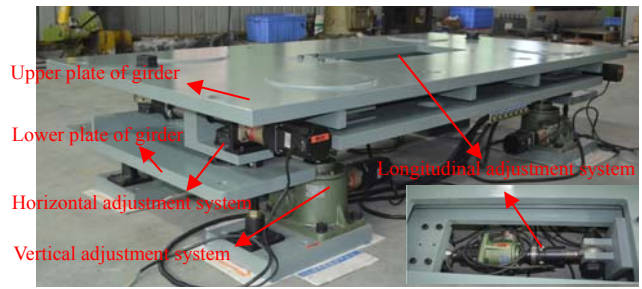

Fig.2. The AC dipole girder structure of CSNS/RCS

The finite element structure (FE) is consists of new dipole, girder system. The dipole magnet is composed of silicon steel sheets, steel plate, stainless plate and aluminum coil; the girder is composed of steel plate and adjusting mechanism. Considering the FE accuracy and computational cost, some measures are used to simplify the FE. The structure physical properties are listed in Table 1.

Table 1. Parameters of material

| Material | Young's modulus (Pa) | Poisson's ratio | Density(Kg.m$^3$) |
|---|---|---|---|
| Q345D steel | 2.09E11 | 0.269 | 7890 |
| Silicon steel sheet | 1.97E11 | 0.26 | 7650 |
| Aluminum | 6.9E10 | 0.34 | 2700 |
| Stainless steel | 1.93E11 | 0.31 | 7750 |

The FE is constructed with element of Solid 186. In the progress of analysis, the Block Lanczos Method is used to calculate the natural frequency and vibration mode of the system. The top 6 step modalities are obtained, and the results are shown in Fig. 3 and Table 2.

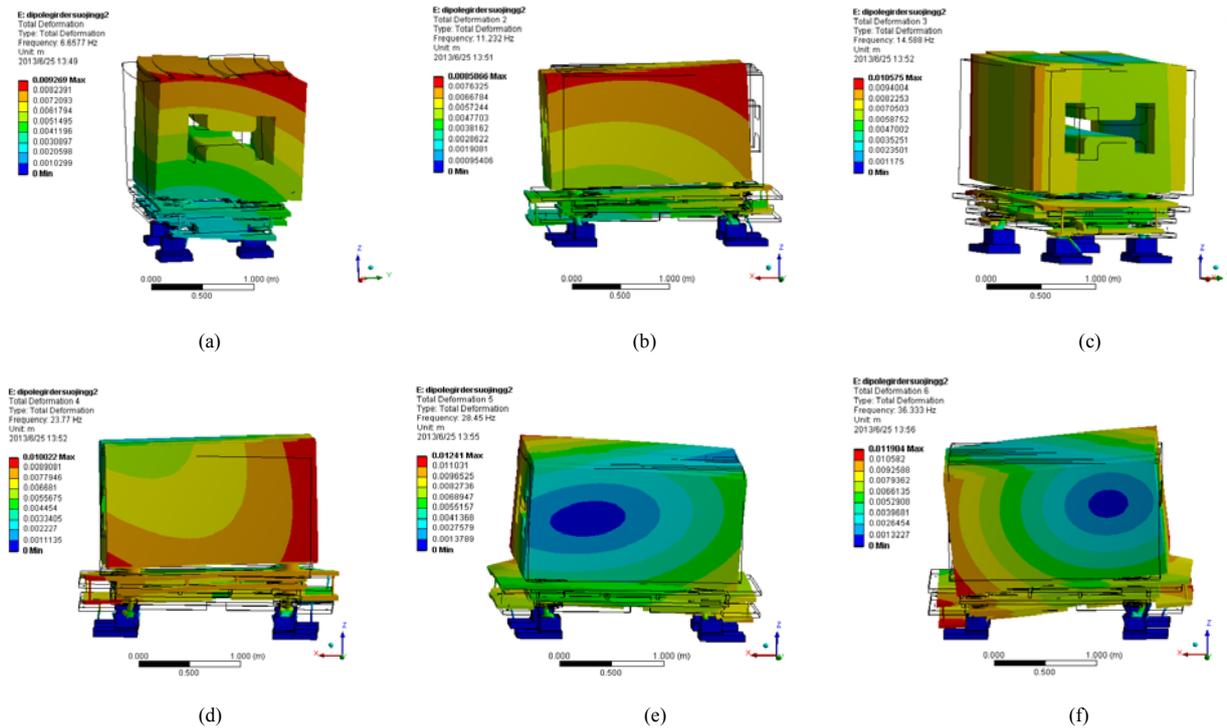

Fig.3. (a), (b), (c), (d), (e), (f) are respectively the top 6 rank modal shapes of the dipole-girder system.

Table 2. The top 6 step natural frequency.

| Modal order i | Natural frequency f/Hz | Modal shape |
|---|---|---|
| 1 | 6.658 | X direction bend |
| 2 | 11.23 | Z direction bend |
| 3 | 14.59 | rotate at Y axis |
| 4 | 23.77 | up-down vertical motion at X-Z plane |
| 5 | 28.45 | pitch motion at one diagonal |
| 6 | 36.33 | pitch motion at the other one diagonal |

### 3.2 Testing modal analysis

Testing modal analysis can obtain the dynamic performance parameters of the system with the curve fitting analyses of the transfer function of the structure's excitation and response (such as acceleration, velocity, displacement, etc.). At the assumption of the zero initial state of system, Eq. (1) is Fourier transformed. And the frequency response function can be obtained based on the orthogonality condition of the real symmetric matrix [7].

$$H_{ij}(\omega) = \sum_{r=1}^{n} \frac{\phi_{ir}\phi_{jr}}{(k_r - \omega^2 m_r) + j\omega c_r} = \sum_{r=1}^{n} \frac{\phi_{ir}\phi_{jr}}{k_r(1 - \lambda_r^3 + j2\zeta_r\lambda_r)}. \quad (3)$$

where $\lambda_r = \omega/\omega_r$, $\omega_r = (k_r/m_r)^{0.5}$, $\zeta_r = c_r/(2 m_r\omega_r)$, $m_r$ is the r step modal mass, $k_r$ and $c_r$ are the r step modal stiffness and modal damping, $\zeta_r$ is the r step modal damping ratio, $\omega_r$ and $\phi_r$ are the r step natural frequency and main modal shape vector. In testing modal, the transfer function can be calculated from the exciting point and detecting point parameters. The different order modal parameters can be calculated from any one row or one column elements.

In this paper, the testing scheme is based on the theoretical modal analysis results of ANSYS. The natural frequency distribution range of the system is estimated. This test takes force hammer excitation system. The modal parameters identification method of MIMO is taken too. There are 60 measuring points arranged around the whole system according to the selecting principle, 24 points arranged on the dipole to measure the X, Y and Z direction acceleration of the 12 corner points, and 36 points arranged on the girder to measure the three direction acceleration of the first and third plate's corner points. The testing system and the acceleration sensor arrangement are shown in Fig. 4.

The force hammer and vibration response signals are acquired by the intelligent analyzer of INV3032C. After testing, the system testing modal parameters (natural frequency and damping ratio) have been got through the data processing analysis system of DASP. The results of the testing are shown in Table 3.

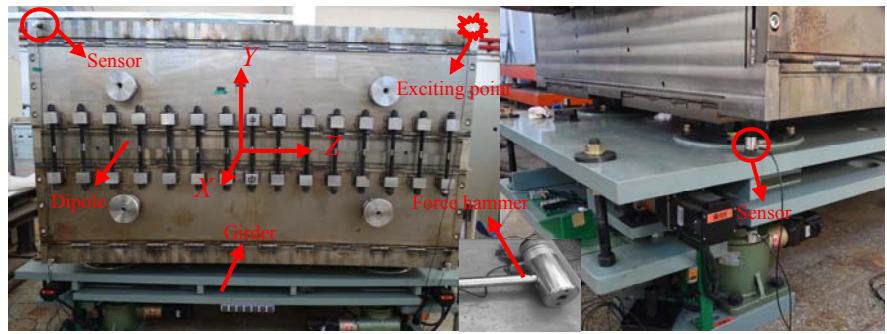

Fig.4. Experiment layout of testing modal

Table 3. The natural frequency and damping ratio of the modal testing

| Modal order | 1 | 2 | 3 | 4 | 5 | 6 |
|---|---|---|---|---|---|---|
| Natural frequency f/Hz | 8.66 | 11.09 | 15.33 | 23.21 | 28.44 | 36.22 |
| Damping ratio /% | 1.772 | 2.602 | 2.647 | 5.519 | 2.786 | 3.922 |

The modal assurance criterion (MAC) is used to estimate the correctness of different mode shape [8].

$$MAC(\{\Psi\}_r, \{\Psi\}_s) = \frac{\left|\{\Psi\}_r^{*T}\{\Psi\}_s\right|^2}{\left(\{\Psi\}_r^{*T}\{\Psi\}_r\right)\left(\{\Psi\}_s^{*T}\{\Psi\}_s\right)}. \quad (4)$$

where $\Psi$ is the mode shape vector. The MAC matrix is one of the important estimate methods in modal parameter identification. The same physical modal MAC value should be close to 1, and the different physical modal MAC value should be very small. Fig.4 is the MAC matrix of the dipole-girder system which satisfies the theory of testing modal analysis. It shows that the experimental modal testing results are correct.

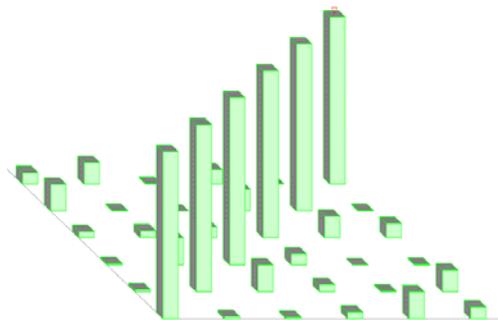

Fig.5. The MAC matrix of modal testing

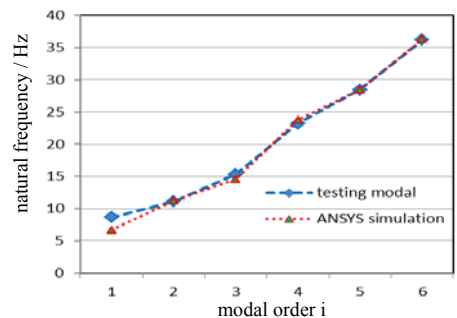

Fig.6. The natural frequencies contrast

Fig 6 shows the natural frequencies contrast between simulations and testing. The simulation results are almost identical with the testing results, which indicates the modal analysis of the structure and the FE of system is reasonable. According to the results, the resonance phenomenon can be avoid at the design stage through ANSYS simulation if the FE is suitable and correct. There isn't a natural frequency closed to the exciting frequency, so system will not amplify the vibration amplitude of the dipole. At the same time the new dipole manufacturing technique was improved. So the new dipole vibration amplitude was decreased to 4.42 um which shows in Fig 7. The dipole vibration can be improved by the suitable dynamic characteristic design.

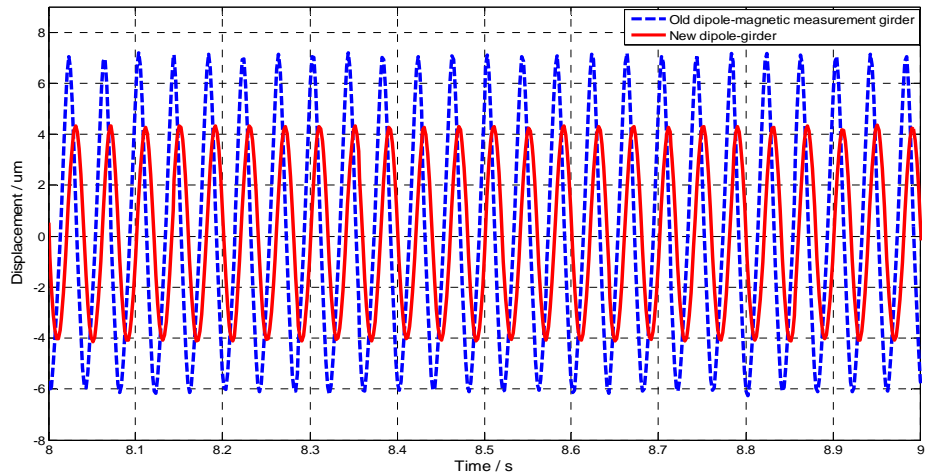

Fig.7. The vibration amplitude contrast of the dipole

## 4 Vibration testing with vibration isolator

The dipole vibratory force still influences the other equipment in the long time operate, although the AC dipole-girder system's dynamic characteristic have been changed and vibration amplitude was less than the old dipole-magnetic measurement girder system. To improve the dynamic characteristic and decrease the dipole vibratory force influence, this paper fix four vibration isolator between the girder and the dipole. The mass of the dipole is 24.5 ton and the centre beam height is 1200 mm. There is only 70mm height for the vibration isolator design which shows in Fig 8. The vibration isolator would have fatigue resistance, creep resistance (<0.1mm per year) and irradiation resistance. The whole factors make the vibration isolator design become a bigger challenge.

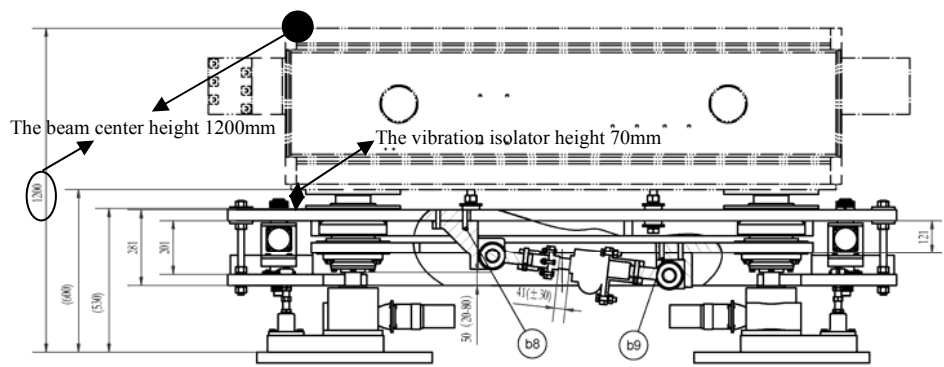

Fig.8. The size of the beam and vibration isolator

In this paper a new absorber was design for the CSNS/RCS AC dipole with special material. Then the acceleration sensor will be used to measure vibration of the dipole, and with vibration isolation testing project are the same testing points as Fig 1. The maximum amplitude is 1.81um at the vertical direction(Y) which shows in Fig. 9.

Submit to "Chinese Physics C"

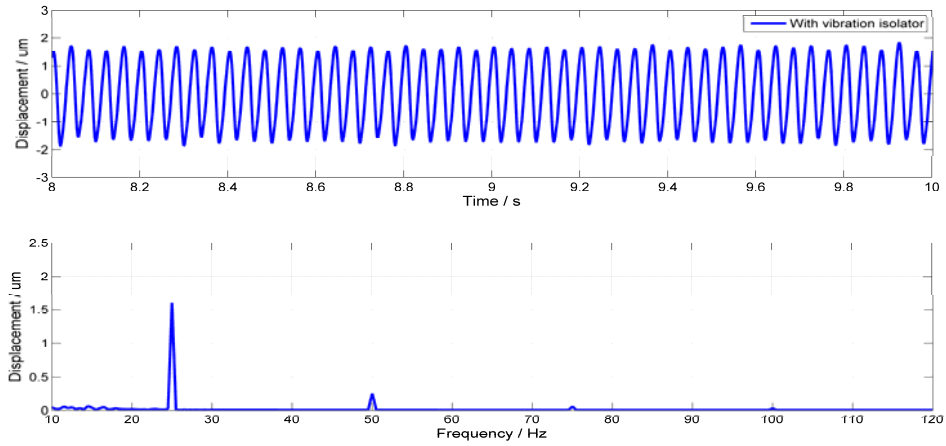

Fig.9. The vibration amplitude (Y direction) with isolator of dipole

From the Fig. 10, the main vibration is not at 25 Hz. The frequency doubling of 50 Hz, 75 Hz and 100 Hz, etc. also have the great contribution to the dipole vibration. The dynamic response of the AC dipole has been changed with vibration isolator. And the exciting vibration (25 Hz) of dipole was reduced, so the dipole vibration amplitude with isolator is decreased 60% compare the vibration amplitude without isolator.

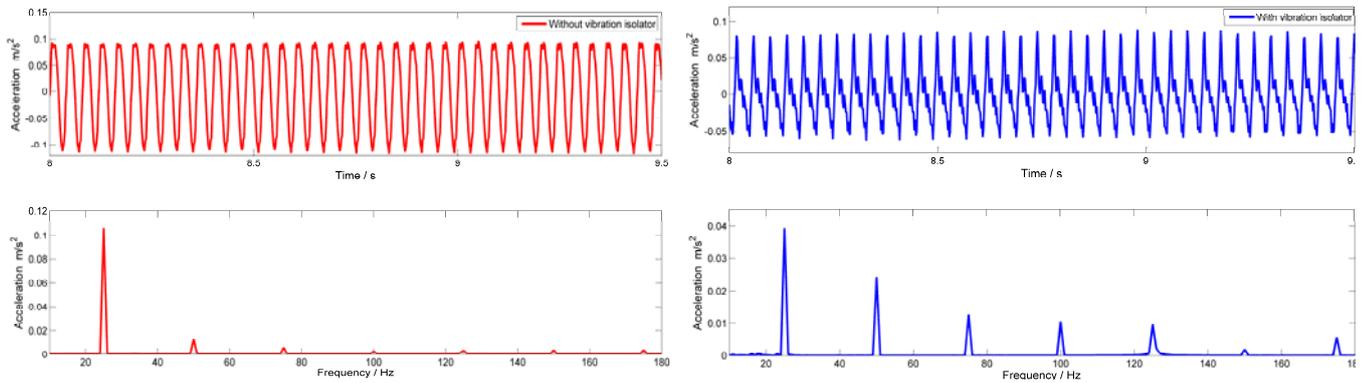

Fig.10. The acceleration amplitude spectrum of dipole(Y direction) comparison of dipole

At the same time the acceleration of the vibration isolator also been measured which used to indicate the vibratory force transmission efficiency decrease. The up sensor acceleration value uses to indicate the vibratory force of the dipole, and the down sensor value uses to indicate the vibratory force after weaken. From the Fig. 11, the up sensor maximum value is 0.17 m/s^2 and the down sensor maximum value is 0.057 m/s^2, so the vibratory force vibration isolation efficiency of the isolator is 66.2%.

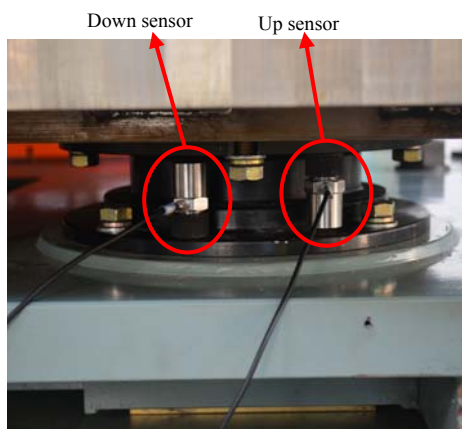 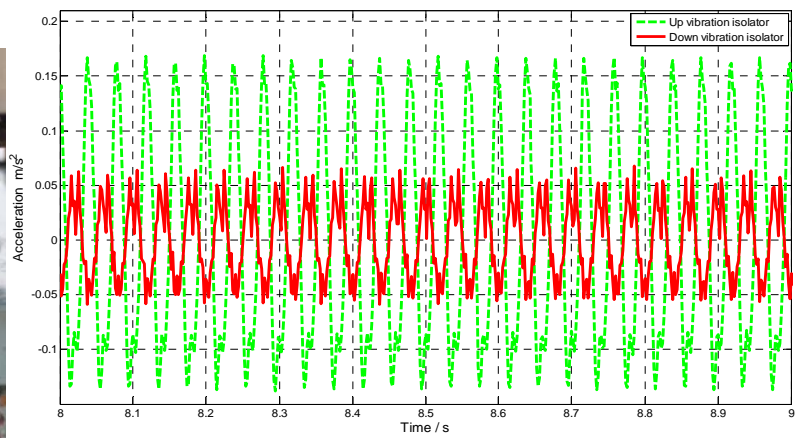

Fig.11. The acceleration comparison between the upside-downside of the vibration isolator

## 5 Conclusions

The AC dipole and girder system play a very important role in the accelerator of CSNS/RCS, so studying the vibration of the system is necessary. This paper established the suitable finite element structure of the magnet girder system, a method for analysing and studying the dynamic characteristic of the system is put forward by combining theoretical calculation (ANSYS simulation) with experimental testing. So the resonance phenomenon can be avoided and the structure vibration resistance can be improved before manufacture. A new isolator was designed to decrease the vibration amplitude and the vibratory force transmission of the AC dipole. With the vibration isolator the AC dipole vibration amplitude decreased 60% and the vibratory force transmission decrease 66.2%. The active vibration of magnet is different with passive vibration which was causes by ground vibration. So this paper can provide a reasonable way to design the equipment which has self-excited vibration such as AC dipole, AC quadrupole [9] and superconductor cavity system of CSNS, etc.

*The authors would like to thank other CSNS colleagues to give help and advice.*